# STATE MACHINE OPERATION OF COMPLEX SYSTEMS*

P. Hanlet†, Fermi National Accelerator Laboratory, Batavia, Illinois, USA

*Abstract*

Operation of complex systems which depend on one, or more, sub-systems with many process variables often operate in more than one state. For each state there may be a variety of parameters of interest, and for each of these, one may require different alarm limits, different archiving needs, and have different critical parameters. Relying on operators to reliably change $10^2$-$10^5$ parameters for each system for each state is unreasonable. Not changing these parameters results in alarms being ignored or disabled, critical changes missed, and/or data archiving inefficiencies.

To reliably manage the operation of complex systems, such as cryomodules (CMs), Fermilab is implementing state machines for each CM and an over-arching state machine for the PIP-II superconducting LINAC (SCL). The state machine transitions and operating parameters are stored/restored to/from a configuration database. Proper implementation of the state machines will not only ensure safe and reliable operation of the CMs, but will help ensure reliable data quality. A description of PIP-II SCL, details of the state machines, and plans for imminent use the state machines for CM testing will be discussed.

## INTRODUCTION

Fermi National Accelerator Laboratory, or Fermilab, in Batavia, Illinois, USA is constructing a new superconducting linear accelerator (LINAC) with twice the energy of the existing LINAC and significantly higher power. The LINAC, PIP-II, will power the rest of the Fermilab accelerator complex, generating the world's most intense high-energy neutrino beam, as well as providing beam to other experiments and test beams, see Fig. 1.

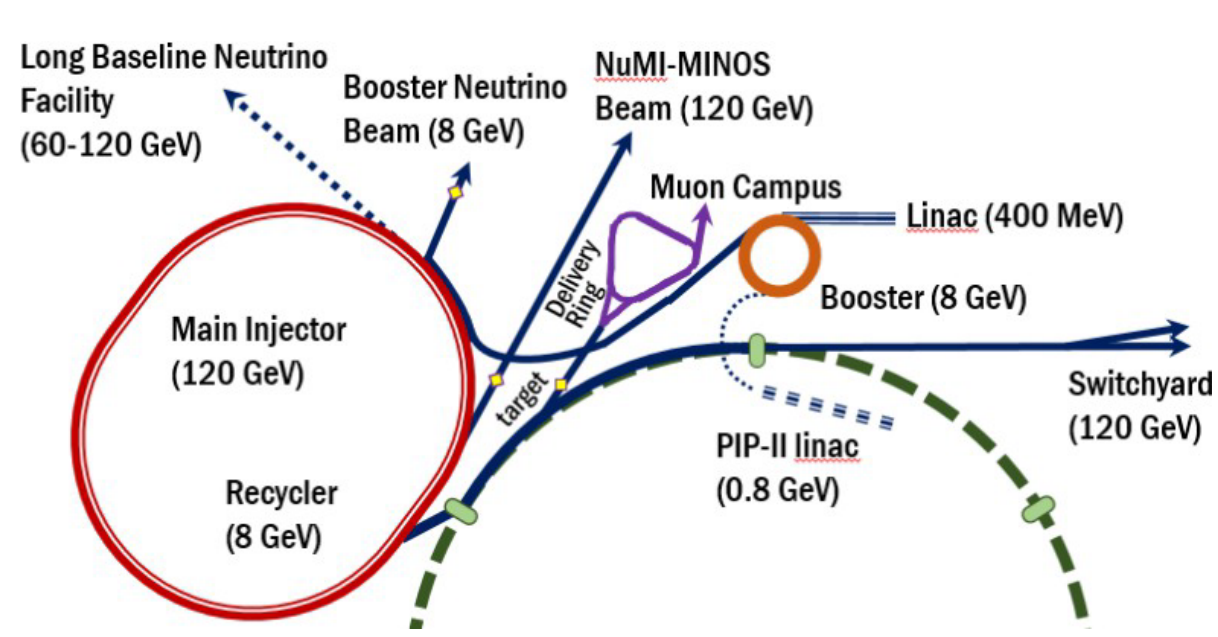

Figure 1: Fermilab's accelerator chain and experiments

The capabilities of PIP-II [1,2] are to provide an 800 MeV proton beam of 1.2 MW using a superconducting RF LINAC, see Fig. 2. The beam is upgradeable to multi-MW and CW-compatible as well as customizable for a variety of user requirements. The scope of PIP-II includes a beam transfer line to the existing Booster ring and accelerator complex upgrades to the Booster, Recycler, and Main Injector,

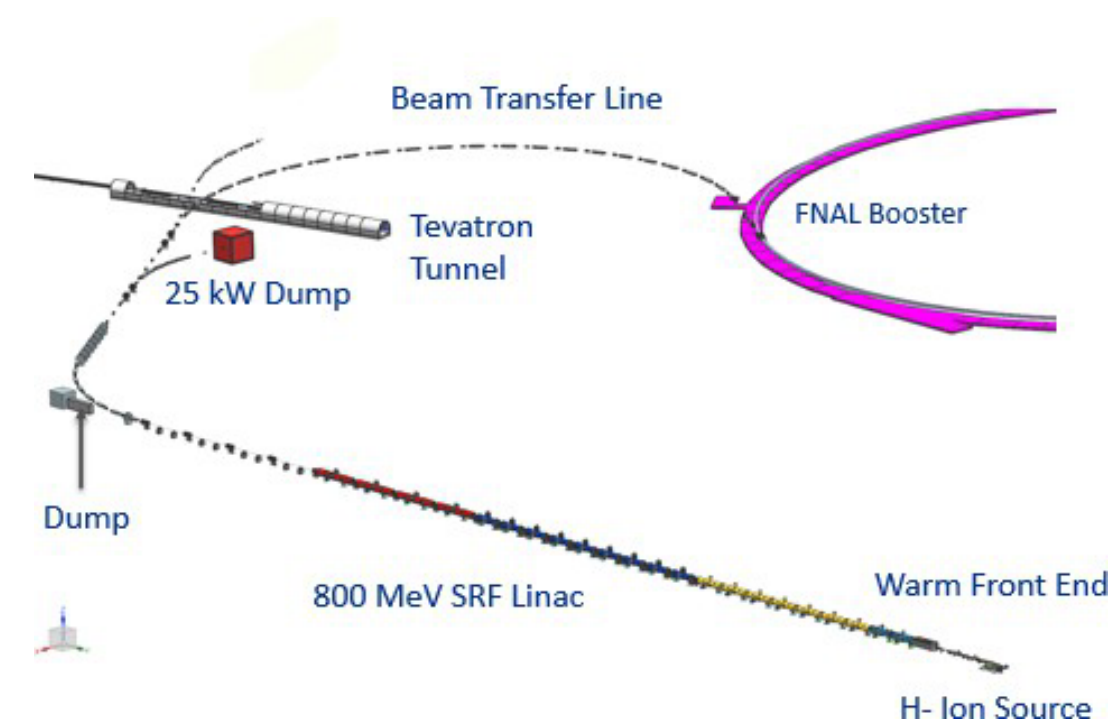

Figure 2: PIP-II scope: new LINAC and beam transfer line to Booster

## MOTIVATION

In this context, a "complex system" refers to a hardware device in a control system which has multiple subsystems and multiple operational states. This frequently means that it has a large number of process variables (PVs) and for each state: there are different PVs of interest, different alarm limits and severities, different archiving needs, and different critical PVs (those used to identify/notify subsystem experts).

The result of ignoring these differences in states can potentially be severe: e.g. incorrect alarm limits may be too loose and thus not notify operators of problems; or worse, if the alarm limits are too tight, the alarms are continuous and likely ignored and/or disabled. For archiving controls data, collection may be inefficient, or worse, changes may not be recorded if dead bands are too loose. Incorrect identification of states may result in delays in operation if subsystem experts are not contacted promptly.

### *Example*

An example of a complex system is an accelerator superconducting RF cryomodule (CM). At PIP-II, each CM will have subsystems: vacuum, cryogenics, safety, machine protection (MPS), RF permits (RFPI), low level RF (LLRF), and high power RF (HPRF). A partial example of these states for this system are shown in Tab. 1. Note that as one transitions through the states, the numbers of PVs change.

Note also the distinction in the "Types" of states. "Static" states refer to those in which all of the PVs are expected to remain constant, within dead bands. Here, the alarm limits are tight and one archives data in a monitor mode. In contrast, for the “Dynamic” states, some of the PVs will remain static, but others are expected to change. Those which change should have a wider alarm limits which span the expected changes, and the data archived in a period mode. The archiving modes will be described the section “Populating Database Tables”.

* This manuscript has been authored by Fermi Research Alliance, LLC under Contract No. DE-AC02-07CH11359 with the U.S. Department of Energy, Office of Science, Office of High Energy Physics.
† hanlet@fnal.gov

Table 1: Example: An incomplete list of finite states for a superconducting RF cryomodule. Here: “ilks" refers to interlocks, “cryo" to cryogenics, and “temps" to temperature.

| State | Static PVs | Dynamic PVs | Type |
|---|---|---|---|
| **Offline** | all | - | static |
| **Pumping** | vacuum: pumps, valves, ilks | gauges | dynamic |
| **Pumped** | vacuum: pumps, valves, ilks, gauges | - | static |
| **HTTS Cooling** | vacuum + cryo: valves, ilks | flow, pressure, temps | dynamic |
| **HTTS Cold** | vacuum + cryo: valves, ilks, flow, pressure, temps | - | static |
| **4K Cooling** | vacuum + more cryo | more flow, pressure, temps, LHe level | dynamic |
| **4K Cold** | vacuum + more cryo | - | static |
| **2K Cooling** | vacuum + more cryo + pump | more flow, temps | dynamic |
| **2K Cold** | vacuum + more cryo + pump, flow | - | static |
| **RF Training** | vacuum + cryo + protection (safety,MPS,RFPI) | LLRF, HPRF | dynamic |
| **Powered** | vacuum + cryo + protection, LLRF, HPRF | - | static |

## STATE MACHINES

Before describing the State Machines (SMs), it is important to recognize what they are *not*. A controls SM is not intended to control the operation of the complex system, in fact, it does not affect the operation of the system in any way. It is certainly not intended for personnel or equipment safety and there are no user interactions with the SM.

### *Description*

A SM is a finite state machine [5]. For our purpose, it is a passive process for a complex system that identifies the PVs which are pertinent to a particular state. The present state is identified by constantly monitoring the PVs that cause a transition to a new state; e.g. from our previous CM example: when certain vacuum valves are open, others are closed, a pump is turned on, and the vacuum gauge reads atmospheric pressure, the CM has transitioned from Offline to Pumping.

### *Algorithm*

For each state then, the SM: (1) identifies the pertinent PVs for the state, (2) collects the PV parameters (alarms limits & severities, archiver dead bands, etc.) for this system and state, (3) dynamically sets these PV fields, (4) adjusts the archiving mode and the pause/unpause request of the PVs, and (5) identifies the critical PVs. This algorithm can be viewed graphically in Fig. 3.

### *Implementation*

The SM is implemented within the EPICS [6] framework invoking the EPICS State Notation Language (SNL). There are three components to the SM:

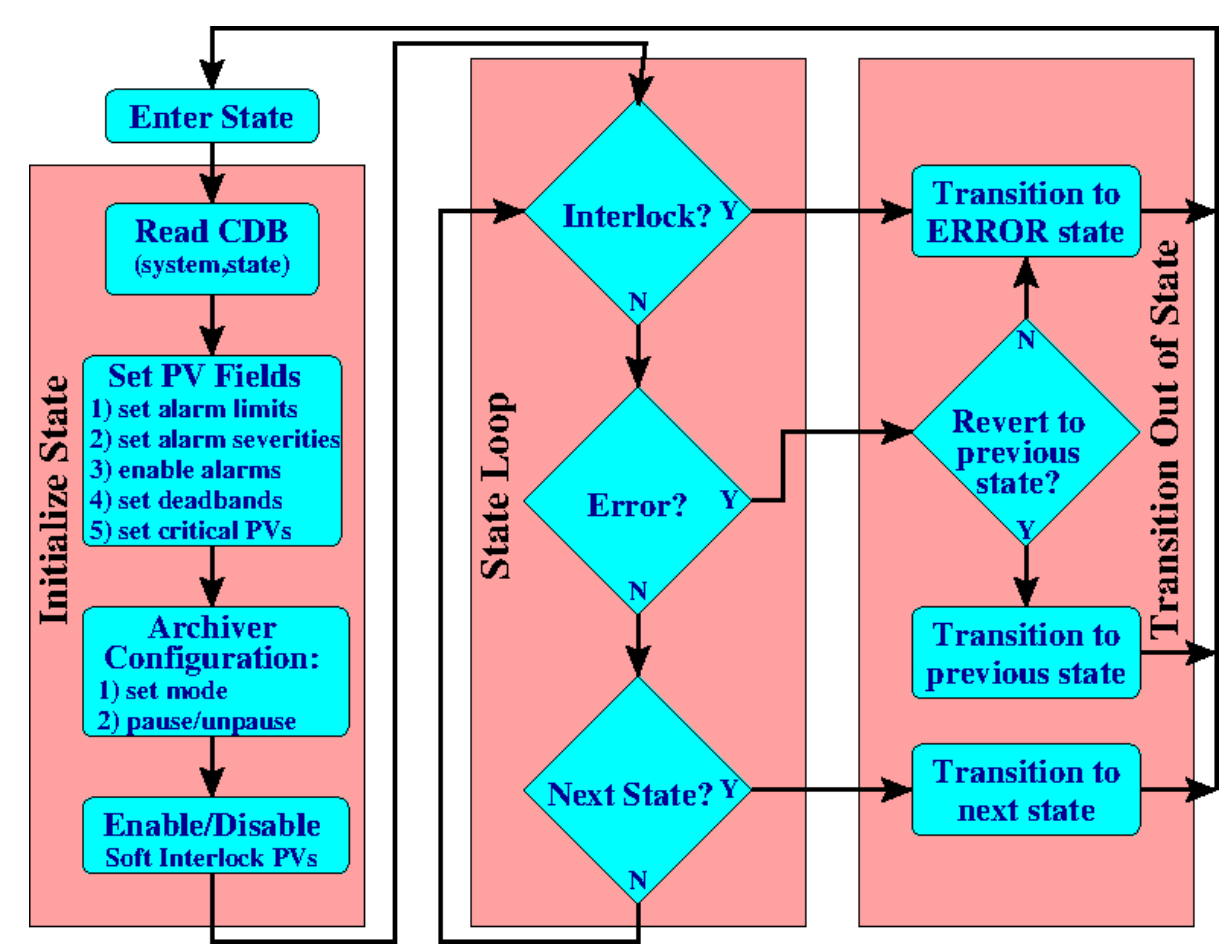

Figure 3: State Machine algorithm: each state of each subsystem follows this sequence.

1. state identifier – a passive program which monitors status or values of PVs, this can be a standalone IOC or part of another controls IOC; this IOC generates the system STATE PV. This function is performed in the “State Loop” and “Transition Out of State” boxes of Fig. 3.

2. configuration database (CDB) – relational database with identical tables (1 table for each state for each complex system). The table holds the state’s pertinent PVs and all of the field values (alarms, etc.) and archiving modes for each PV.

3. a sequencer IOC implemented in EPICS SNL which:

    - monitors system STATE PV
    - reads database table associated with new state
    - dynamically changes PV alarm limits, severities, dead bands
    - dynamically changes Archiver functionality

- starts a monitor of critical PVs,

where the "critical PVs" are associated with subsystem experts and used to notify them when their system exhibits problems. The third step is performed in the "Initialize State" box of Fig. 3. The variables read from the CDB which are used to set the fields for each record in the new state are: LOLO, LOW, HIGH, HIHI, LLSV, LSV, HSV, HHSV, and ADEL. The archiver mode is changed to/from "Monitor" or "Scan", and PVs archiving may be unpaused or paused.

### *Populating Database Tables*

The most difficult task in the implementation of the SM is populating the CDB tables. For PIP-II, this is being performed by the subsystem experts. They are provided a template spreadsheet and taught how to populate it. A snippet of such a spreadsheet is shown in Fig. 4. In the top-left corner of the figure are the identifiers for the complex system and the state, corresponding to a single table in the CDB. Here, one can also see the column headers; these are identical to the headers in the CDB tables. Each row, in Fig. 4 is a PV. The color-coded groupings belong to different subsystems. There is one sheet for each state for the complex system and each sheet may have input from multiple subsystem owners.

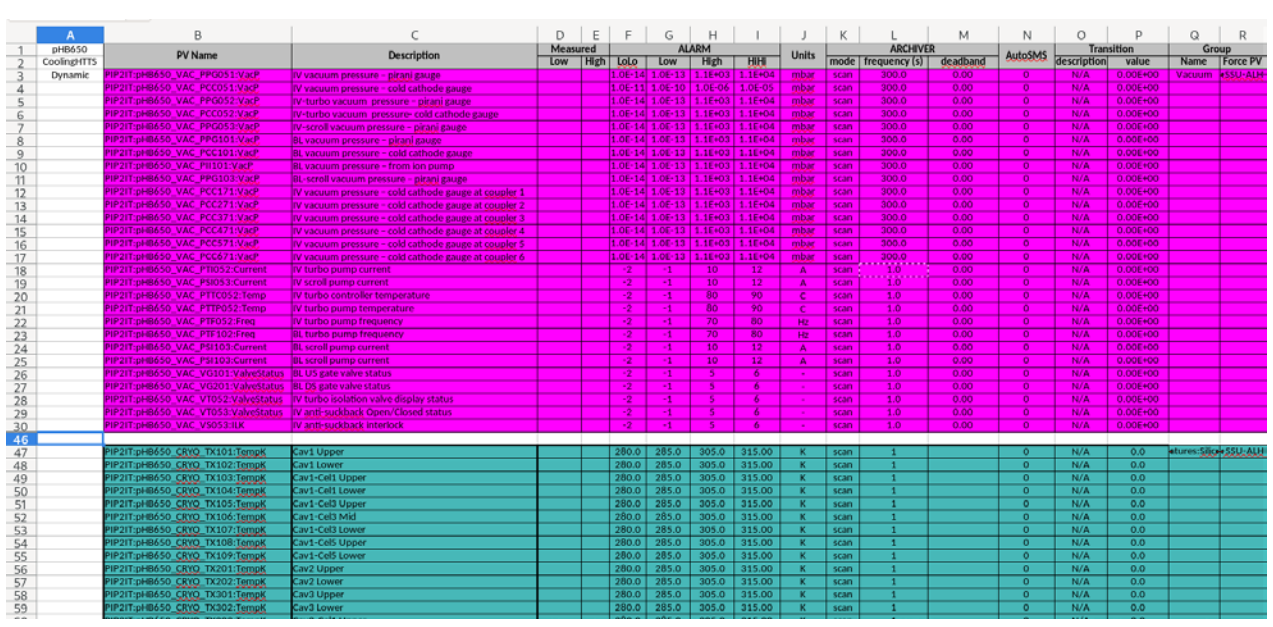

Figure 4: Snippet of an SM spreadsheet for populating CDB tables

Grouping PVs in the spreadsheet simplifies the process, and if existing, either archived data or simulation predictions can help set ranges for alarm limits and dead bands. A combination of logbook entries and archive data can help define the start and end of a state. As an example of how we populate the SM spreadsheets, consider Fig. 5, which shows a time plot cooldown of the 6 RF couplers at the 5K and 50K intercepts of the prototype HB650 cryomodule at the PIP2IT CM test stand. One can easily see features: initial cooldown, cryo plant hiccough, a thermalization period, and additional cooling when the cavities, and consequently couplers, were being cooled. Referring to Tab. 1, these states correspond to HTTS Cooling, HTTS Cold, 4K Cooling. From this plot, we identify the alarm limits for the dynamic HTTS Cooling state. Furthermore, for the static HTTS Cold state, we calculate the mean and standard deviation for each coupler sensor and define temperature alarm limits as LOLO=$<T> -3\sigma$, LOW=$<T>-2\sigma$, HIGH=$<T>+2\sigma$, and HIHI=$<T>+3\sigma$.

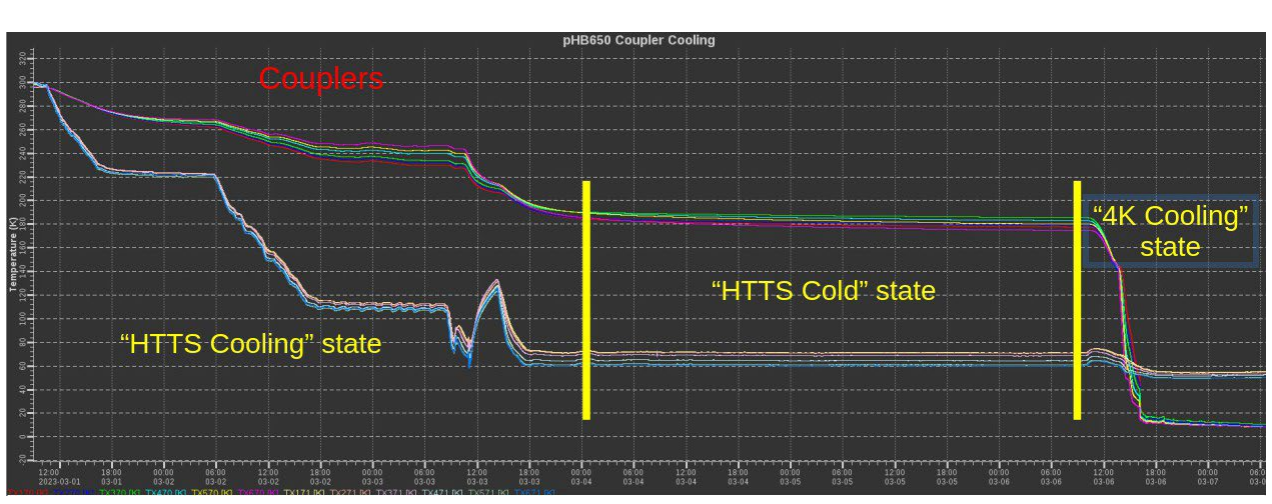

Figure 5: Archived data of the RF coupler cooldown of a prototype HB650 CM

For archiving, we scan the data periodically for the dynamic HTTS Cooling state. For the static HTTS Cold state, we continue to scan at 1 Hz, but archive the data in monitor mode with a dead bands of $<T>\pm1\sigma$. In this way, one scans that values as before, but reduces electronic noise while still being sensitive to any events which cause a significant change.

For other PVs it may be more difficult to establish these parameters. Note that setting alarm limits and dead bands is frequently an iterative process. Once the spreadsheet is populated, we use a python script to load the values into the CDB for production.

## TESTING AT PIP2IT

PIP2IT is the PIP-II Integration Test Stand and is in the Cryomodule Test Facility (CMTF) at Fermilab. This facility will be used to test PIP-II cryomodules prior to installing them in the LINAC.

From 2021 to June 2023, much of the control system infrastructure has been established for PIP2IT: EPICS IOCs were developed and services were installed. The services include HMIs (using Phoebus), Archiver Appliance, Alarm Handling, Channel Finder, and Save and Restore. Significant work remains so that the services are correctly configured and robust. The goal is that these systems be vetted and made robust, as well as to train operators, prior to establishing a reliable control system at PIP-II.

Two phases of testing have already taken place on the prototype HB650 CM. When it returns to PIP2IT in early 2024, it will operate with its state machine.

## OPERATION AT PIP-II

As previously stated, there is no user interaction with a SM, though it is conceivable that there be states that are set manually; e.g. a "Testing" state, in which one operates with interlocks bypassed. Nonetheless, there is value in displaying the state of a complex system.

Figure 6 shows an early prototype the main operator display which may be used for PIP-II operations. The purpose of the page is to be both an accelerator status at a glance and a launcher for each complex system. For each CM, the status of each subsystem is displayed in stop-light style below it. Similarly, for beamline accelerator components and beamline instrumentation, a stop-light status is also displayed above the image of the PIP-II LINAC.

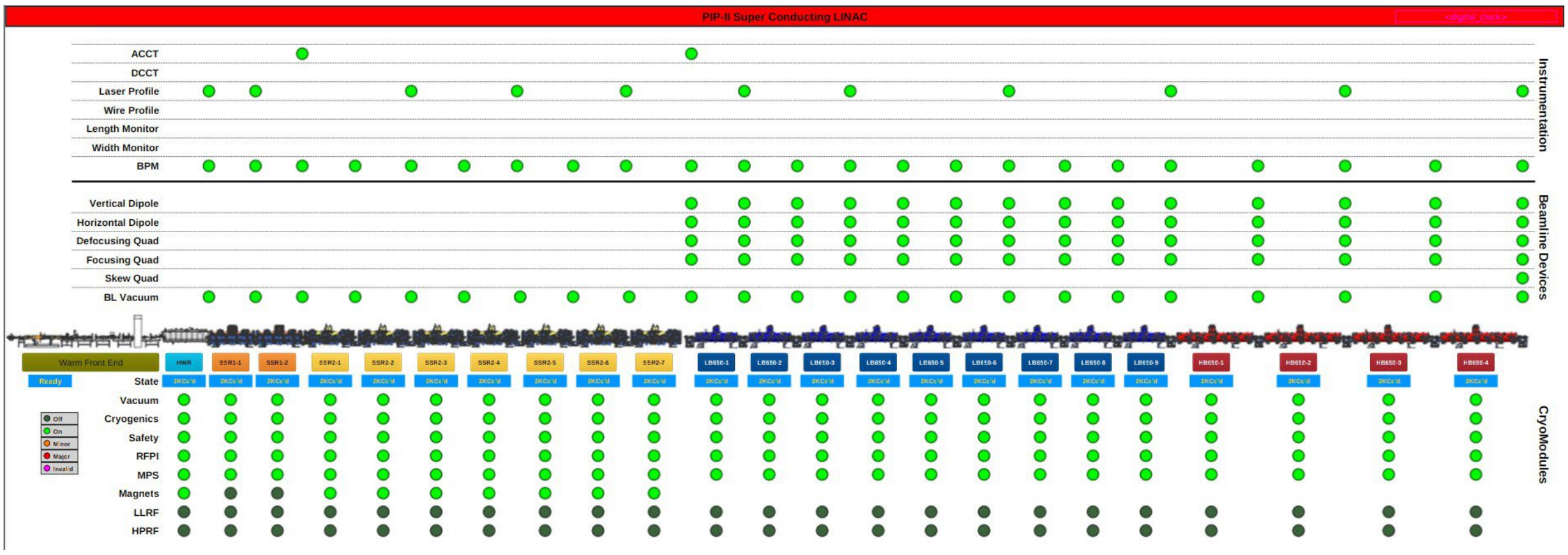

Figure 6: Prototype operator main page for PIP-II accelerator [1].

An expanded view of Fig. 6 is shown in Fig. 7. Here one can see the STATE PVs displayed from the SMs for both SSR2-7 and LB650-1.

Explicitly displaying the STATE PVs for each system allows operators to quickly assess the readiness of major systems for operation.

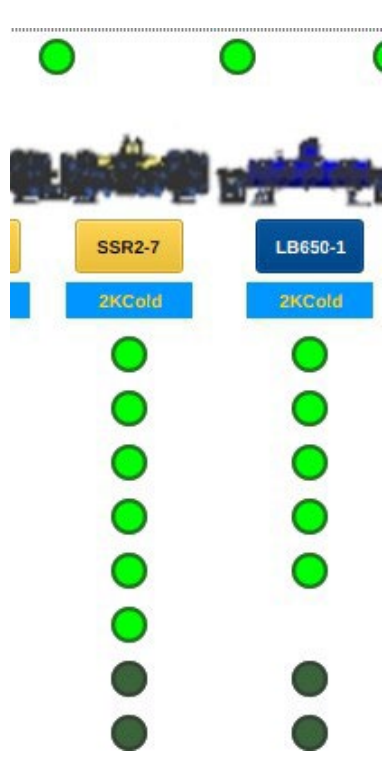

Figure 7: Zoom in on Fig. 6 displaying the STATE PVs for SSR2-7 and LB650-1.

## OTHER TARGETS

Within the context of PIP-II, there are other target complex systems which are under consideration for operation with SMs. The warm front end of the PIP-II LINAC is a likely candidate, as is the cryogenics plant. Consideration is also being given to the beamline accelerator components and beamline instrumentation.

ACORN [7], which is responsible for modernizing the control system for the remainder of the Fermilab accelerator complex, is presently in the process of evaluating options for the future control system and has expressed interest in SMs.

## CONCLUSIONS

Fermilab's flagship project, PIP-II, will provide a 1.2 MW, 800 MeV, CW-compatible, proton beam to the LBNF target to create the world's most intense beam of neutrinos to DUNE, as well as beam for a variety of experiments at Fermilab for decades to come.

A State Machine is a powerful tool to be used to assure that operation of the superconducting RF LINAC is robust with appropriate alarms and archiving for the room temperature proton source and the 33 cryomodules. PIP2IT will use the state machine during testing of the cryomodules at CMTF and allow us to both vet the state machine procedures for each flavor of cryomodules, as well as provide its services during testing.

## ACKNOWLEDGEMENTS

This manuscript has been authored by Fermi Research Alliance, LLC under Contract No. DE-AC02-07CH11359 with the U.S. Department of Energy, Office of Science, Office of High Energy Physics.